\RequirePackage{fix-cm}
\documentclass[smallextended]{svjour3} 

\usepackage{mathptmx}       
\usepackage{helvet}         
\usepackage{courier}        
\usepackage{type1cm}        
\usepackage{makeidx}         
\usepackage{graphicx}        
\usepackage{multicol}        
\usepackage[bottom]{footmisc}

\usepackage{amsmath}

\usepackage{tabularx}
\usepackage{colortbl}

\usepackage[table]{xcolor}
\definecolor{forestgreen}{HTML}{228B22}
\definecolor{indigo}{HTML}{4B0082}

\begin{document}

\title{Evolution of Threats in the Global Risk Network}
\author{Xiang Niu$^{1,2+}$ \and Alaa Moussawi$^{1,3}$ \and Gyorgy Korniss$^{1,3}$ \and Boleslaw K. Szymanski$^{1,2+*}$}
\institute{Boleslaw K. Szymanski \at $^1$Network Science and Technology Center, Rensselaer Polytechnic Institute (RPI), Troy, NY 12180, USA.\\$^*$Correspondence: szymab@rpi.edu\\$^+$Equal contributors}
%
%
\maketitle
\newcolumntype{C}[1]{>{\centering\arraybackslash} m{#1}}

\begin{abstract}
With a steadily growing population and rapid advancements in technology, the global economy is increasing in size and complexity. This growth exacerbates global vulnerabilities and may lead to unforeseen consequences such as global pandemics fueled by air travel, cyberspace attacks, and cascading failures caused by the weakest link in a supply chain. Hence, a quantitative understanding of the mechanisms driving global network vulnerabilities is urgently needed. Developing methods for efficiently monitoring evolution of the global economy is essential to such understanding. Each year the World Economic Forum publishes an authoritative report on the state of the global economy and identifies risks that are likely to be active, impactful or contagious.
Using a Cascading Alternating Renewal Process approach to model the dynamics of the global risk network, we are able to answer critical questions regarding the evolution of this network. To fully trace the evolution of the network we analyze the 
asymptotic state of risks (risk levels which would be reached in the long term if the risks were left unabated) given a snapshot in time; this elucidates the various challenges faced by the world community at each point in time. We also investigate the influence exerted by each risk on others. Results presented here are obtained through either quantitative analysis or computational simulations.
\end{abstract}

\section*{Introduction}
Recently, cascading failures have been extensively studied, with most studies on infrastructure systems~\cite{moussawi2017limits,dobson2007complex}, financial institutions~\cite{haldane2011systemic,gai2010contagion,battiston2012debtrank}, and the Internet~\cite{oppenheimer2003internet,brown2001embracing}. 
In~\cite{moussawi2017limits}, the authors analyze cascading failures in power grid networks. In real and synthetic spatial systems, they study properties of failures and strategies to reduce the corresponding damages. In either single-node or multi-nodes cases, the damage caused by attacks is weakly correlated with network properties, such as node degree or initial state load. They also test different mitigation strategies and various combinations of node failures. Knowing in advance a set of failing nodes and damage incurred by every single-node failure, they are able to predict the damage caused by sets of multi-node failures. That is because, in the multi-node failure, the node with the highest damage dominates the entire multi-node cascading failure. Authors in~\cite{dobson2007complex} also focus on blackout cascading failure mechanisms. Studying the real blackouts from a few countries, they find that frequencies of blackouts exhibit a power-law distribution in agreement with the scale-free property of complex networks. The authors' power system model suggests that the real power system gradually reaches a critical point.

In the area of financial institutions, authors in~\cite{haldane2011systemic} discuss the failures of banking ecosystems. Inspired by models from food webs and disease networks, the authors apply an analogous model in financial networks to reduce risks. In~\cite{gai2010contagion}, the authors study the contagion in the financial market. They mainly focus on the robustness of asset market liquidity. With a novel model based on the Poisson random graph in the banking system, they find that high average degree of nodes increases both the probability of transmission and the speed of contagion. Using the feedback-centrality, the authors in~\cite{battiston2012debtrank} propose a DebtRank to find the critical nodes that play the most important role in the systemic failure of financial networks. They studied a real dataset from Fed emergency loans program and detected 22 institutions that were critical to 2008-2010 crises. They also find that some small institutions can be important because of their high centralities in the network. 
Unlike here, the authors do not even attempt to match the model with historical data. In~\cite{kenett2015partial,kenett2012dependency,raddant2016interconnectedness,havlin2015cascading}, the authors observe historical correlations between stock prices and build stock dependency networks. They also consider events that may affect the stock market such as "tsunami in Japan". In their node failure analysis, they mainly focus on the failure of one industry and test the system tolerance of such failure. In summary, the research in the above papers targets the cascade failures in the global financial market.

Some other analyses are related to the cascade failure of the Internet. In~\cite{oppenheimer2003internet}, the authors analyze the fault-tolerance of Internet service. They establish that Internet failures are mainly caused by operator errors and suggest the use of extensive online testing to reduce the failure rates. In~\cite{brown2001embracing}, the authors find that the latent errors are likely to accumulate within the Internet services and cause chain reaction cascades.  In~\cite{majdandzic2014spontaneous}, the authors analyze spontaneous recovery from cascading failures of economy. In their model, a node can fail independently or by external causes. All nodes have the same internal failure probability so the authors can solve the model using mean-field equation.

Only a few of studies discussed above focus on the cascades of global risks~\cite{szymanski2015failure,lin2017limits}. Yet, the global risks impact highly global economy and lives of countless people. Hence, there is an urgent need to study and understand global risk network.
Here, we model such network using a Cascading Alternating Renewal Processes (CARP)~\cite{szymanski2015failure,lin2017limits,cox1977theory}. In the model, a system alternates between active and passive states, denoted by 1 and 0 respectively.
An active risk represents a failed node, while a passive risk corresponds to a fully operational node. State transitions are instantaneous. They are triggered by non-homogeneous Poisson processes~\cite{szymanski2015failure}. Given the complexity of real-world network interactions and node specific dynamics, the processes causing state transitions may be observable, or latent. The latent processes are not directly observable, only their combined effect, a state transition, is. In the global risk model, the latent processes are categorized as endogenous (caused internally at the node) or exogenous (caused by neighboring active risk node) Poisson processes. Their parameters are recovered by maximum likelihood estimation from the records of historical events.

The risks listed in the WEF Global Risk Reports~\cite{WEF2013,WEF2014,WEF2015,WEF2016,WEF2017} constantly change; new risks arise and are added to the network, while existing active risks either continue to be a threat and remain in the network, or, thanks to the response of threatened governments and industry, decline in importance and are removed. This evolution causes continuous changes in the global risks and their probabilities, and leads to an annual revision of the list of risks present in the network. However, if left unabated, the global risk network would approach the steady state, in which some risk will be active much more frequently making their threats much more pronounced than in the initial state. These considerations motivate us not to compare the states of the global risk network at fixed points in time. Instead, we compare the steady states to which these initial states would evolve if no changes to the system had been introduced. Looking at the steady states accentuates the different challenges that the well-being and stability of the global risks network has faced at each reported point in the time.

The CARP model has been successfully applied to analyze cascading failures of global risks in~\cite{szymanski2015failure} by taking into account the interconnectivity and interdependence among risks. This model is also used here to analyze the evolution of the global risk network.
Furthermore, in~\cite{lin2017limits}, the authors investigate the asymptotic normality of the MLE procedure used to find the most likely model parameters in CARP. They demonstrate that this property is preserved in the presence of latent processes causing state transitions. We use this property here to bind the error of model parameter recovery in the global risk network.

In the following sections of the paper we analyze evolution of risks over the years 2013-2017. Some of these results were presented in~\cite{niu2017evolution} but they were limited to two points in time, year 2013 and year 2017. 
Hence, the number of the results and points of evolution presented here more than doubled. Consequently, all the analyses and evaluations were expanded accordingly. In particular, we first present the annual evolution of the risk network itself from 2013 to 2017. We also show how the definitions of risks themselves change and how we deal with these changes. Finally, we present and discuss how steady states of risks evolve annually.

\section*{Models}
\subsection*{CARP}
The CARP model for the global risk network contains two primary types of Poisson processes: two latent passive risk activation processes and a directly observable active risk continuation process. Here, we further subdivide the passive risk activation type into internal activation and external activation. We assume that the active risk continuation process is always triggered internally. The corresponding Poisson processes are defined as follows~\cite{szymanski2015failure}.
\begin{itemize}
	\item Passive risk with internal activation: a passive risk $i$ is activated internally with intensity $\lambda_i^{int}$. The Poisson probability of transition over one time unit is $p_i^{int}=1-e^{-\lambda_i^{int}}$.
	\item Passive risk with external activation: a passive risk $i$ is activated externally by the neighboring active risk $j$ with intensity $\lambda_{ji}^{ext}$. The corresponding Poisson probability is $p_{ji}^{ext}=1-e^{-\lambda_{ji}^{ext}}$.
	\item Active risk continuation: an active risk $i$ continues its activity for the next time unit internally with intensity $\lambda_{i}^{con}$. The corresponding Poisson probability is $p_{i}^{con}=1-e^{-\lambda_{i}^{con}}=1-p_{i}^{rec}$, where $p_i^{rec}$ denotes probability of recovery in a time unit from an active risk $i$.
\end{itemize}

Using the likelihood $l_i$ for each risk $i$ provided by experts in the WEF Global Risk Reports~\cite{WEF2013,WEF2014,WEF2015,WEF2016,WEF2017}, we obtain a normalized likelihood $L_i$, which indicates how likely a risk $i$ is to be active by logarithmic transformation $\lambda_i^{int}=-\alpha\ln(1-L_i)$, $\lambda_{ji}^{ext}=-\beta\ln(1-L_i)$, $\lambda_i^{con}=-\gamma\ln(1-L_i)$~\cite{szymanski2015failure} getting:
\begin{eqnarray}
p_i^{int}&=&1-(1-L_i)^{\alpha} \nonumber\\
p_{ji}^{ext}&=&1-(1-L_i)^{\beta} \nonumber\\
p_i^{con}&=&1-(1-L_i)^{\gamma}  .
\label{eq_poisson}
\end{eqnarray}
The advantage of Eq.~\ref{eq_poisson} is that the probabilities of the three Poisson processes are defined only by a normalized likelihood $L_i$ and model parameters $\alpha, \beta, \gamma$. These parameters are needed because while humans can often adequately estimate relative probabilities by crowd sourcing, they usually are less precise in predicting absolute probabilities. By providing a mapping from likelihood to probabilities based on the most likely values of the model parameters obtained through MLE procedure~\cite{pawitan2001all} they model over historical data, we account for expert biases that may render the absolute probabilities inaccurate, while extracting relevant information from the relative estimations of the likelihoods.

After combining the probabilities of all possible transitions of the Poisson processes in the risk network, we obtain the state transition probabilities~\cite{szymanski2015failure} as
\begin{eqnarray}
P_i(t)^{0\rightarrow1}&=&1-(1-p_i^{int})\prod_{j\in N_i} (1-p_{ji}^{ext}) \nonumber\\
P_i(t)^{1\rightarrow0}&=&p_i^{rec}.
\label{eq_transition}
\end{eqnarray}
where $P_i(t)^{s_1\rightarrow s_2}$ represents the probability of a state transition from state $s_1$ to state $s_2$ for risk $i$ at time $t$, while $N_i$ represents the set of neighbors of risk $i$ that were active at time $t-1$.

The specific state transition probability of all risks over one time unit is $\overrightarrow{S(t)}=\prod_{i=1}^{R} P_i(t)^{s_i(t)\rightarrow s_i(t+1)}$, where we sum over all $R$ risks. The likelihood of the sequence of state transition of all risks is $L(\overrightarrow{S(1)},\overrightarrow{S(2)},...,\overrightarrow{S(T)})=\prod_{t=1}^{T-1} \prod_{i=1}^{R} P_i(t)^{s_i(t)\rightarrow s_i(t+1)}$, where $T$ represents the number of time steps during the entire observed or predicted evolution~\cite{szymanski2015failure}. The corresponding log-likelihood is 
\begin{eqnarray}
\ln L(\overrightarrow{S(1)},\overrightarrow{S(2)},...,\overrightarrow{S(T)})=\sum_{t=1}^{T-1} \sum_{i=1}^{R} \ln P_i(t)^{s_i(t)\rightarrow s_i(t+1)}  .
\label{eq_mle}
\end{eqnarray}

We compute the most likely values of model parameters $\alpha, \beta$, and $\gamma$ by maximizing the log-likelihood (Eq.~\ref{eq_mle}) of the observed state transitions over the historical data~\cite{dempster1977maximum,pawitan2001all}. With these most likely values, we simulate the system evolution to identify the mean-field steady state evolution of the global risk network at any particular point in time. Given the state of the global risk network at time $t$ we ask to what state the network will evolve as $t \rightarrow \infty$ if it is not further influenced by external actors. We simulate the system evolution as $t \rightarrow \infty$ training on historical data till time $t$, and relying only on learned model dynamics past time $t$. We approximate the asymptotic state at a finite time at which activity frequencies for all risks stabilize.

\begin{table}
\caption{Indices and descriptions of global risks from the year 2013 to 2017. To compare related risks in different years, we give them same numerical code but different alphabetical indices here. In the World Economic Forum (WEF) Global Risk Reports, experts define risks in 5 categories: economic (blue), environmental (green), geopolitical (orange), societal (red) and technological (purple).}
\centering
\scriptsize
\begin{tabular}{|c|c|c|c|c|c|c|} \hline
\cellcolor{gray} \shortstack{Risk\\index} & \cellcolor{gray} Risk description & \cellcolor{gray} \shortstack{2013\\index} & \cellcolor{gray} \shortstack{2014\\index}  & \cellcolor{gray} \shortstack{2015\\index}  & \cellcolor{gray} \shortstack{2016\\index}  & \cellcolor{gray} \shortstack{2017\\index}  \\ \hline
{\color{blue}01} & {\color{blue}Fiscal crises in key economies} & {\color{blue}01} & {\color{blue}01} & {\color{blue}06} & {\color{blue}05} & {\color{blue}05} \\ \hline
{\color{blue}} & {\color{blue}High structural unemployment} & {\color{blue}} & {\color{blue}} & {\color{blue}} & {\color{blue}} & {\color{blue}} \\
{\color{blue}02} & {\color{blue}or underemployment} & {\color{blue}02} & {\color{blue}04} & {\color{blue}07} & {\color{blue}06} & {\color{blue}06} \\ \hline
{\color{blue}} & {\color{blue}Failure of a major financial mechanism} & {\color{blue}} & {\color{blue}} & {\color{blue}} & {\color{blue}} & {\color{blue}} \\ 
{\color{blue}03} & {\color{blue}or institution} & {\color{blue}05} & {\color{blue}02} & {\color{blue}04} & {\color{blue}03} & {\color{blue}03} \\ \hline
{\color{blue}04} & {\color{blue}Failure/shortfall of critical infrastructure} & {\color{blue}06} & {\color{blue}06} & {\color{blue}05} & {\color{blue}04} & {\color{blue}04} \\ \hline
{\color{blue}05a} & {\color{blue}Severe energy price shock} & - & {\color{blue}05} & {\color{blue}03} & {\color{blue}08} & {\color{blue}08} \\ \hline
{\color{blue}05b} & {\color{blue}Extreme volatility in energy and agriculture prices} & {\color{blue}03} & - & - & - & - \\ \hline
{\color{blue}06a} & {\color{blue}Asset bubble in a major economy} & - & - & {\color{blue}01} & {\color{blue}01} & {\color{blue}01} \\ \hline
{\color{blue}06b} & {\color{blue}Liquidity crises} & {\color{blue}07} & {\color{blue}03} & - & - & - \\ \hline
{\color{blue}07a} & {\color{blue}Deflation in a major economy} & - & - & {\color{blue}02} & {\color{blue}02} & {\color{blue}02} \\ \hline
{\color{blue}07b} & {\color{blue}Unmanageable inflation} & - & - & {\color{blue}08} & {\color{blue}09} & {\color{blue}09} \\ \hline
{\color{blue}07c} & {\color{blue}Unmanageable inflation or deflation} & {\color{blue}10} & - & - & - & - \\ \hline
{\color{blue}} & {\color{blue}Decline of importance of the US dollar} &  & {\color{blue}} & & & \\ 
{\color{blue}07d} & {\color{blue}as a major currency} & - & {\color{blue}07} & - & - & - \\ \hline
08 & Severe income disparity & {\color{blue}08} & {\color{red}25} & - & - & - \\ \hline
{\color{blue}09} & {\color{blue}Unforeseen negative consequences of regulation} & {\color{blue}09} & - & - & - & - \\ \hline
{\color{blue}10} & {\color{blue}Hard landing of an emerging economy} & {\color{blue}04} & - & - & - & - \\ \hline
{\color{forestgreen}11} & {\color{forestgreen}Extreme weather events} & {\color{forestgreen}16} & {\color{forestgreen}08} & {\color{forestgreen}09} & {\color{forestgreen}10} & {\color{forestgreen}10} \\ \hline
{\color{forestgreen}} & {\color{forestgreen}Failure of climate-change mitigation} & {\color{forestgreen}} & {\color{forestgreen}} & {\color{forestgreen}} & {\color{forestgreen}} & {\color{forestgreen}} \\ 
{\color{forestgreen}12} & {\color{forestgreen}and adaptation} & {\color{forestgreen}12} & {\color{forestgreen}13} & {\color{forestgreen}10} & {\color{forestgreen}11} & {\color{forestgreen}11} \\ \hline
{\color{forestgreen}13} & {\color{forestgreen}Major biodiversity loss and ecosystem collapse} & {\color{forestgreen}18} & {\color{forestgreen}11} & {\color{forestgreen}11} & {\color{forestgreen}12} & {\color{forestgreen}12} \\ \hline
{\color{forestgreen}14a} & {\color{forestgreen}Major natural catastrophes} & - & {\color{forestgreen}09} & {\color{forestgreen}12} & {\color{forestgreen}13} & {\color{forestgreen}13} \\ \hline
{\color{forestgreen}14b} & {\color{forestgreen}Unprecedented geophysical destruction} & {\color{forestgreen}19} & - & - & - & - \\ \hline
{\color{forestgreen}14c} & {\color{forestgreen}Vulnerability to geomagnetic storms} & {\color{forestgreen}20} & - & - & - & - \\ \hline
{\color{forestgreen}15a} & {\color{forestgreen}Man-made environmental catastrophes} & - & {\color{forestgreen}10} & {\color{forestgreen}13} & {\color{forestgreen}14} & {\color{forestgreen}14} \\ \hline
{\color{forestgreen}15b} & {\color{forestgreen}Irremediable pollution} & {\color{forestgreen}13} & - & - & - & - \\ \hline
{\color{forestgreen}15c} & {\color{forestgreen}Land and waterway use mismanagement} & {\color{forestgreen}14} & - & - & - & - \\ \hline
{\color{forestgreen}15d} & {\color{forestgreen}Rising greenhouse gas emissions} & {\color{forestgreen}17} & - & - & - & - \\ \hline
16 & Antibiotic-resistant bacteria & {\color{forestgreen}11} & {\color{red}26} & - & - & - \\ \hline
{\color{orange}17} & {\color{orange}State collapse or crisis} & {\color{orange}21} & {\color{orange}15} & {\color{orange}17} & {\color{orange}18} & {\color{orange}19} \\ \hline
{\color{orange}18} & {\color{orange}Weapons of mass destruction} & {\color{orange}22} & {\color{orange}19} & {\color{orange}18} & {\color{orange}19} & {\color{orange}20} \\ \hline
{\color{orange}19} & {\color{orange}Interstate conflict with regional consequences} & {\color{orange}24} & {\color{orange}20} & {\color{orange}15} & {\color{orange}16} & {\color{orange}17} \\ \hline
{\color{orange}20} & {\color{orange}Large-scale terrorist attacks} & {\color{orange}28} & {\color{orange}18} & {\color{orange}16} & {\color{orange}17} & {\color{orange}18} \\ \hline
21a & Illicit trade & {\color{orange}30} & - & - & {\color{blue}07} & {\color{blue}07} \\ \hline
{\color{orange}21b} & {\color{orange}Entrenched organized crime} & {\color{orange}23} & - & - & - & - \\ \hline
{\color{orange}} & {\color{orange}Major escalation in organized crime} & & {\color{orange}} & & & \\ 
{\color{orange}21c} & {\color{orange}and illicit trade} & - & {\color{orange}17} & - & - & - \\ \hline
{\color{orange}22a} & {\color{orange}Failure of national governance} & - & - & {\color{orange}14} & {\color{orange}15} & {\color{orange}15} \\ \hline
{\color{orange}22b} & {\color{orange}Pervasive entrenched corruption} & {\color{orange}27} & {\color{orange}16} & - & - & - \\ \hline
{\color{orange}23} & {\color{orange}Failure of global governance} & {\color{orange}25} & {\color{orange}14} & - & - & {\color{orange}16} \\ \hline
{\color{orange}24} & {\color{orange}Unilateral resource nationalization} & {\color{orange}29} & {\color{orange}21} & - & - & - \\ \hline
{\color{orange}25} & {\color{orange}Militarization of space} & {\color{orange}26} & - & - & - & - \\ \hline
26 & Failure of urban planning & {\color{forestgreen}15} & {\color{red}27} & {\color{red}19} & {\color{red}20} & {\color{red}21} \\ \hline
{\color{red}27} & {\color{red}Food crises} & {\color{red}32} & {\color{red}22} & {\color{red}20} & {\color{red}21} & {\color{red}22} \\ \hline
28 & Water crises & {\color{red}40} & {\color{forestgreen}12} & {\color{red}24} & {\color{red}25} & {\color{red}26} \\ \hline
{\color{red}29a} & {\color{red}Rapid and massive spread of infectious diseases} & - & - & {\color{red}23} & {\color{red}24} & {\color{red}25} \\ \hline
{\color{red}29b} & {\color{red}Rising rates of chronic disease} & {\color{red}35} & {\color{red}24} & - & - & - \\ \hline
{\color{red}29c} & {\color{red}Vulnerability to pandemics} & {\color{red}39} & {\color{red}23} & - & - & - \\ \hline
{\color{red}30} & {\color{red}Large-scale involuntary migration} & {\color{red}37} & - & {\color{red}21} & {\color{red}22} & {\color{red}23} \\ \hline
{\color{red}31} & {\color{red}Profound political and social instability} & - & {\color{red}28} & {\color{red}22} & {\color{red}23} & {\color{red}24} \\ \hline
{\color{red}32} & {\color{red}Backlash against globalization} & {\color{red}31} & - & - & - & - \\ \hline
{\color{red}33} & {\color{red}Ineffective illicit drug policies} & {\color{red}33} & - & - & - & - \\ \hline
{\color{red}34} & {\color{red}Mismanagement of population aging} & {\color{red}34} & - & - & - & - \\ \hline
{\color{red}35} & {\color{red}Rising religious fanaticism} & {\color{red}36} & - & - & - & - \\ \hline
{\color{red}36} & {\color{red}Unsustainable population growth} & {\color{red}38} & - & - & - & - \\ \hline
{\color{indigo}} & {\color{indigo}Breakdown of critical information infrastructure} & {\color{indigo}} & {\color{indigo}} & {\color{indigo}} & {\color{indigo}} & {\color{indigo}} \\ 
{\color{indigo}37} & {\color{indigo}and networks} & {\color{indigo}41} & {\color{indigo}29} & {\color{indigo}25} & {\color{indigo}27} & {\color{indigo}28} \\ \hline
{\color{indigo}38} & {\color{indigo}Large-scale cyberattacks} & {\color{indigo}42} & {\color{indigo}30} & {\color{indigo}26} & {\color{indigo}28} & {\color{indigo}29} \\ \hline
{\color{indigo}39} & {\color{indigo}Massive incident of data fraud/theft} & {\color{indigo}45} & {\color{indigo}31} & {\color{indigo}27} & {\color{indigo}29} & {\color{indigo}30} \\ \hline
{\color{indigo}40a} & {\color{indigo}Adverse consequences of technological advances} & - & - & {\color{indigo}28} & {\color{indigo}26} & {\color{indigo}27} \\ \hline
{\color{indigo}40b} & {\color{indigo}Massive digital misinformation} & {\color{indigo}44} & - & - & - & - \\ \hline
{\color{indigo}40c} & {\color{indigo}Proliferation of orbital debris} & {\color{indigo}47} & - & - & - & - \\ \hline
{\color{indigo}} & {\color{indigo}Unforeseen consequences of climate change} & {\color{indigo}} & & & & \\ 
{\color{indigo}40d} & {\color{indigo}mitigation} & {\color{indigo}48} & - & - & - & - \\ \hline
{\color{indigo}40e} & {\color{indigo}Unforeseen consequences of nanotechnology} & {\color{indigo}49} & - & - & - & - \\ \hline
{\color{indigo}} & {\color{indigo}Unforeseen consequences of new life science} &  & & & & \\ 
{\color{indigo}40f} & {\color{indigo}technologies} & {\color{indigo}50} & - & - & - & - \\ \hline
{\color{indigo}41} & {\color{indigo}Failure of intellectual property regime} & {\color{indigo}43} & - & - & - & - \\ \hline
{\color{indigo}42} & {\color{indigo}Mineral resource supply vulnerability} & {\color{indigo}46} & - & - & - & - \\ \hline
\end{tabular}
\label{table_risk_index}
\end{table}

\subsection*{Related Works}
The CARP model for global risk network was first proposed and analyzed in detail in~\cite{szymanski2015failure}. Through the study of 2013 network, the authors calculate the contagion potentials of risks, risk persistence and risk failure cascade survival probability caused by a single risk failure. The risk persistence is calculated as the fraction of time steps during which a risk is active. The contagion potential is not positively correlated with internal activation probability, but is mainly defined by external activation and recovery probabilities. Ranked by their contagion potentials, the top three risks are: "Severe income disparity", "Chronic fiscal imbalances", and "Rising greenhouse gas emissions". The results show that about 80\% of the time, the number of active risks is between 8 and 19. By setting internal activation probabilities of risks to zero, the authors found cascade survival probability initiated by a single risk decreases exponentially with time. To validate the choice of the model, the authors compared it with 60 alternative models, including disconnected model ($\beta=0$, so only internal activation is acting), expert data based model ($\alpha=1, \beta=0$, internal activation is equal to likelihood $L_i$), uniform model (likelihood $L_i$ is ignored), weighted network model (edges are assigned different weights depending on the number of experts listing them) and combinations of them. The CARP model outperforms all other models  with at least 95\% statistical confidence interval because it takes into account the interconnectivity and interdependence among risks.

The precision of predictability is likely to depend on the quality and amount of historical ground truth data. Thus, in~\cite{lin2017limits}, the authors proposed an artificial model of fire propagation among houses to establish the limits of model predictability. The authors use simulations of CARP model to generate data with arbitrary lengths (from 100 to 6400 time steps) and arbitrary number of variants of model execution. The authors use these variants as alternative historical ground truth data. They measure the prediction precision between variants at different length of alternative historical data, and study how the prediction precision changes over time. The authors conclude that the average relative error of parameter recovery decays according to the power law of the size of historical data and ultimately tends to zero when the length of historical data tends to infinity. These results demonstrate that the asymptotic normality of MLE holds also in the presence of latent Poisson processes. 

There are some similarities between the CARP model~\cite{cox1977theory} and epidemic models, such as SIS~\cite{epidemic}, if we consider risks as a population undergoing infection with the activation pathogen. Yet deeper comparison reveals that the CARP model is more complex by including latent exogenous (becoming sick by infection) and directly observable endogenous (becoming sick without or not through contact with infected nodes) activation. Thus, finding model parameters matching historical data is more complex in CARP model than in epidemic model. Another significant difference is the  evolution of transition probabilities and risk population, as new threats arise, old ones die, and some existing risks change their probability to activate as a result of the increasing resilience developed by threatened governments, organizations, and people.


\begin{figure}
	\centering
	\includegraphics[width=0.8\textwidth]{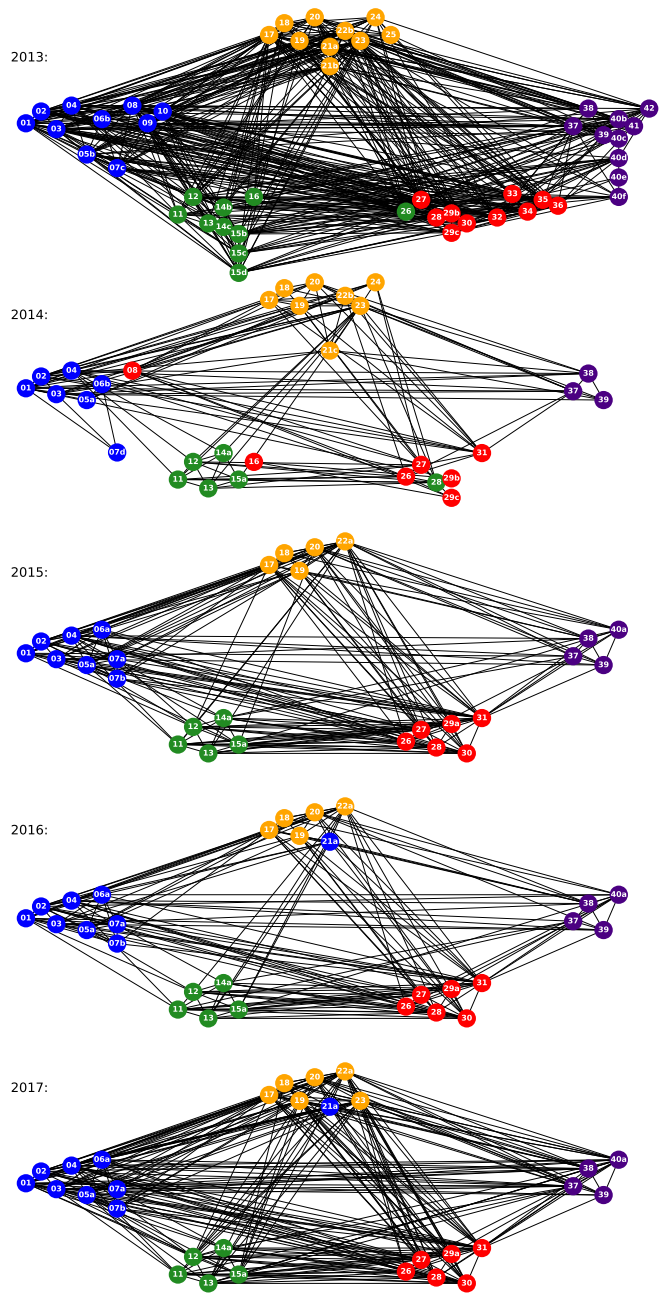}
	\caption{The evolution of risk networks over five years from 2013 to 2017. Although the number of risks vary over the years, the five groups of risk remain unchanged. We use the same layout for each year network and fix the relative positions of risks with the same index over the years. The risks are in five color groups and five location groups. A color group is the category of a risk in a certain year labeled by experts, while locations of the groups are labeled by us and remain unchanged over the years. 
The groups are as follows: economic, with mostly blue nodes placed leftmost of the network and including risks 01-10; environmental, with mostly green nodes placed the second to the left and at the bottom part of the figure with risk 11-16; geopolitical, with mostly orange nodes placed the middle and at the top of the figure with risk 17-25; societal, with mostly red nodes placed the second to the right and at the bottom of the figure with risk 26-36; technological, with purple nodes placed at the rightmost of the figure with risk 37-42.}
	\label{fig_network_evolution}
\end{figure}

\begin{figure}
	\centering
	\includegraphics[width=\textwidth]{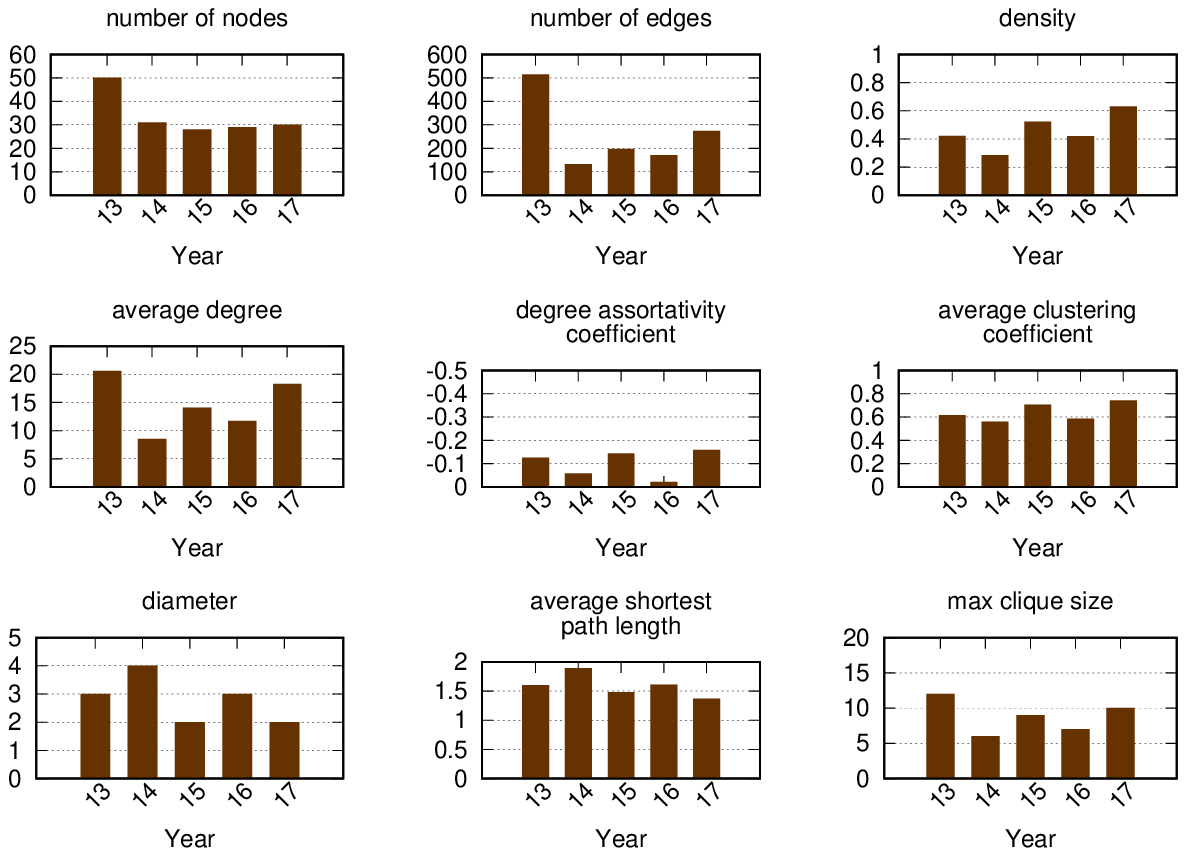}
	\caption{The evolving risk network properties over five years from 2013 to 2017. The 2013 network is the largest and also contains most edges. However, it is not the one with the highest density, since 2015 and 2017 networks are much smaller and with a large number of edges. Generally, 2013, 2015 and 2017 networks have relatively high average degrees, degree assortativity coefficients, average clustering coefficients, and max clique sizes and relatively low diameters and average shortest path lengths.}
	\label{fig_network_property}
\end{figure}

\section*{Risk Network Evolution}

In the World Economic Forum (WEF) Global Risk Reports~\cite{WEF2013,WEF2014,WEF2015,WEF2016,WEF2017}, experts define risks in five categories: economic, environmental, geopolitical, societal and technological. The list of risks is shown in Table~\ref{table_risk_index}. We use five different colors to differentiate between risk categories and to aid the understanding of the WEF network, because each year the risks are categorized slightly differently. To track related risks, we give them identical numerical codes. The categories of risks vary over the years as well. For example, risk 21a ``Illicit trade" is in the geopolitical category in 2013 and 2014 when experts felt illicit activity and crime were probable risks. In 2016 and 2017 this risk is categorized as economic, considering that such trade impacts more the global economy than geopolitical factors.
In the 2013 risk network, the risks are uniformly distributed over the five categories. From 2014 to 2017, the economic risks category is the largest and contains around eight risks; the environmental, geopolitical and societal categories contain approximately six risks, and the technological category is the smallest with mostly four risks.

Fig.~\ref{fig_network_evolution} and~\ref{fig_network_property} show how the global risk network and it's properties have changed over time. In Fig.~\ref{fig_network_evolution}, each node represents different risk from Table~\ref{table_risk_index}. Each undirected unweighted link of two endpoints represents that the two risks are related in opinion of some WEF experts. For the WEF report, each of the experts was asked to answer the question: "Global risks are not isolated and it is important to assess their interconnections. In your view, which are the most strongly connected global risks? Please select three to six pairs of global risks." Then the interconnection $w_{ij}$ between risks $i$ and $j$ is calculated as:
\begin{equation}
\begin{split}
w_{ij} &= \sqrt{\frac{\sum_{n=1}^{N}pair_{ij,n}}{pair_{max}}}\\
pair_{max} &= \max_{ij}(\sum_{n=1}^N pair_{ij,n})
\end{split}
\end{equation}
where $pair_{ij,n}$ is 1 if the risks $i$ and $j$ are interconnected from the perspective of expert $n$, otherwise it is 0~\cite{WEF2017}. In paper~\cite{szymanski2015failure}, the authors report on testing whether the model with weighted edges outperforms the one used here which has unweighted edges and current model was statistically significantly better than the weighted edge alternative. By definition, these edges represent risk relationships, thus they act as transmission channels for risk propagation through the external activation process.
The 2013 risk network is the largest, with the greatest average degree. Despite this, due to the smaller size of the 2017 network, its risks have greater interconnectivity, a larger mean clustering coefficient, and a smaller diameter. Most subfigures in Fig.~\ref{fig_network_property} show that the 2013, 2015, and 2017 networks are denser than the 2014 and 2016 networks.

\section*{Historical Events}

We utilize and update the event dataset created for~\cite{szymanski2015failure}, which included news, academic articles, Wikipedia entries, etc. from Jan. 2000 to Dec. 2012, and from which we collected 13x12x50=7,800 data points for the 2013 risk network. For the 2014 to the 2017 risk networks, we relabel prior events and collect new events dated from Jan. 2013 to Dec. 2016. Thus the total number of data points is now 17x12x62=12,648, for 62 risks in Table~\ref{table_risk_index}. Each data point indicates if a risk is active or passive in a certain month. By maximizing the log-likelihood defined by Eq.~\ref{eq_mle} for the observed state transitions~\cite{dempster1977maximum,pawitan2001all}, we obtain the most likely values of model parameters $\alpha, \beta, \gamma$ for each year.

\begin{figure}
	\centering
	\includegraphics[width=\textwidth]{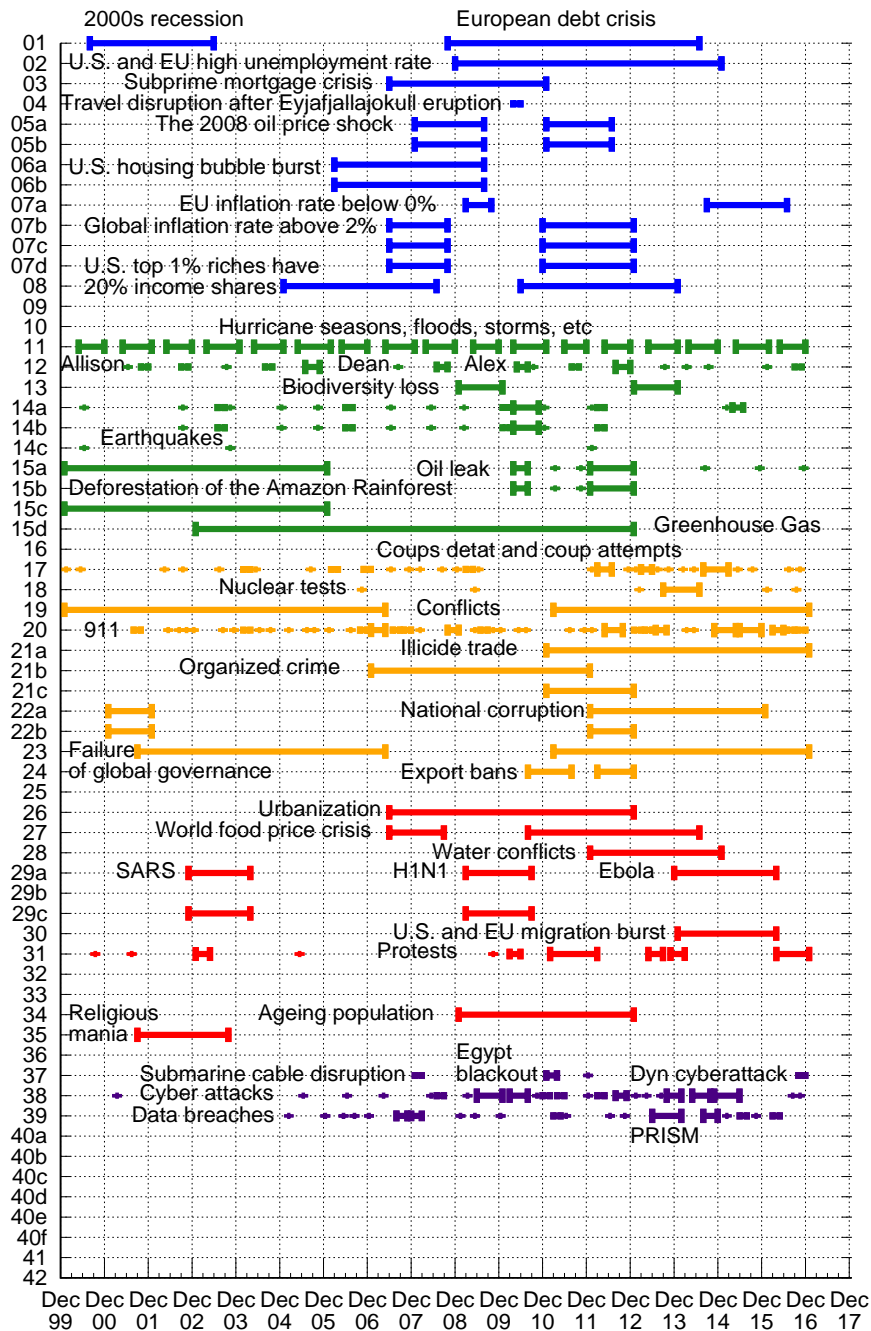}
	\caption{The historical risk events from 2000 to 2017. We searched thousands of events online over 18 years and selected hundreds from them as risk-related events. We label the events based on the description of risks in each year WEF Global Risk Report. The events are also grouped into five categories.}
	\label{fig_historical_events}
\end{figure}

Fig.~\ref{fig_historical_events} shows the timeline of historical events. Among economic risks, activation of some events is recorded directly based on the corresponding Wikipedia articles, such as ``European debt crisis", ``Subprime mortgage crisis" and ``Air travel disruption after the 2010 Eyjafjallajokull eruption". Activation of other events is identified through human processing of the online statistical data. ``U.S. high unemployment rate" is recorded when the United States (U.S.) unemployment rate is above 7.5\%, ``EU high unemployment rate" is recognized when the European Union (EU) unemployment rate is above 9.5\%. The ``Oil price shock" is recognized when the yearly change of oil price per barrel exceeds \$40 which happened during 2008 and 2009, and during 2011 and 2012. ``U.S. housing bubble burst" is recorded when the average new house purchase price in the U.S., reaches above \$280,000. ``EU deflation" is recognized when the inflation rate is below 0\% in the EU, ``Global inflation" is recorded when major economic regions such as the U.S. and the EU have the inflation rate above 2\%. ``U.S. severe income disparity" is recognized when U.S. top 1\% of U.S. richest people own more than 20\% of total incomes. Most of the economic risk events activated at around 2008 but then became passive after 2014. There is also a chain reaction among them, ``U.S. housing bubble" caused ``Subprime mortgage crisis", then led to ``U.S. and EU high unemployment rate" and ``European debt crisis".

Among environmental risks, ``Hurricane seasons" is identified according to the yearly Atlantic and Pacific hurricane seasons, which happen regularly every year, starting from May and ending in December. Risk 12 ``Failure of climate-change mitigation and adaptation" is recognized when there are tremendous damages caused by climate change such as hurricanes Allison, Dean, and Alex. We consider damage as tremendous when the cost is above \$1 billion. Risk 13 ``loss of biodiversity" is recognized when common bird index drops below 100. Risk 14 ``Major natural catastrophes" are identified by deadliest earthquakes, avalanches, wildfires, heat waves, solar storms, etc. Those events also happen naturally, but last shorter than extreme weather events. ``Deforestation of the Amazon Rainforest" is recorded when yearly deforestation rate is above 15,000 $km^2$ which happened from 2000 to 2005. ``Deepwater Horizon oil spill" happened in 2010, ``Beijing air pollution soars to hazard level" was observed in 2012. Unlike economic risk events, those events usually last up to a year. ``Rising greenhouse gas emissions" is recorded when Annual Greenhouse Gas Index (AGGI) exceeds 1.2.

Among geopolitical risks, Risk 17 ``State collapse or crisis" is recognized by all coups d'\'etat and coup attempts. Most of them ends within a month. Among few events related to Risk 18 ``Weapons of mass destruction", we list ``Destruction of Syria's chemical weapons" in 2013 and several ``North Korean nuclear test" started in 2006. Risk 19 ``Interstate conflict" activation list includes major global conflicts from 2000: ``War on Terror", ``Second Congo War", ``Syrian Civil War", ``Iraqi Civil War", and ``Cold War II". The ``War on Terror" was triggered by ``911 attack" in 2001 and includes the wars in Afghanistan, Iraqi, Syria, etc. We consider it ended in 2007 when British government abandoned the use of the term. But the conflicts between nations and terrorists continue to happen. The ``Second Congo War" from 1998 to 2003 involved nine nations and often is referred to as ``Africa's World War". The ``Syrian Civil War" from 2011 and the ``Iraqi Civil War" from 2014 are sometimes described as ``proto-world war". ``Cold War II" refers to the political tension between two opposing geopolitical sides, with one led by Russia and China, and the other led by the United States and NATO. It starts at the same time with the ``Ukraine crisis" event activation in 2013. The events related to Risk 20 ``Large-scale terrorist attack" are selected from the worldwide worst terrorist strikes each of which caused at least 300 injuries, 100 fatalities or 20 fatalities among children. Illicit financial flows from developing countries are above \$1 trillion per year from 2011. We consider activation of Risk 22b ``Pervasive entrenched corruption" occurs when there is at least one country in the world with the Corruption Perceptions Index below 10\% (small index represents high corruption). The scope of Risk 23 ``Failure of global governance" is close to Risk 19 ``Interstate conflict", since it includes the inability to resolve issues of terrorism, wars, political and economic tensions between countries. The geopolitical risks are closely related and share similar activity scope: 2001-2006 and 2012-2016. We can see correlation between events related to those risks. These events started with ``911" followed by the increase of the number and damages of terrorist attacks.

Among the societal risks, there were two major world food price crises during 2007-2008 and 2010-2014, flagged by the rise of the FAO deflated food price index above 150. The peak of the number of ``water conflicts" was reached during 2012-2014 (above 15 conflicts per year) in the Middle East, which is most likely caused by the geopolitical risks. Risk 29a ``Rapid and massive spread of infectious diseases" includes infectious diseases that cause more than thousands death and had worldwide impact, such as ``SARS", ``H1N1" and ``Ebola". The new disease occurs every several years and it also last takes several years before it stops spreading. Risk 30 ``Large-scale involuntary migration" includes migration crises in the U.S. and the EU. In ``U.S. migration crisis", tens of thousands of women and children from El Salvador, Guatemala, and Honduras migrated to the United States in 2014. In ``EU migration crisis", more than 50,000 refugees were arriving in EU each month in 2015. The refugees mainly come from Eritrea, Nigeria, Somalia, Syria, and Afghanistan. Risk 31 ``Profound political and social instability" includes major global protests. Risk 34 ``Mismanagement of population ageing" mostly activates in developing countries such as China and India. Risk 35 ``Rising religious fanaticism" mostly materializes in the Middle East.

Among technological risks, Risk 37 ``Breakdown of critical information infrastructure and networks" materializes every several years. But each time it lasts a short period. The source of the risk could be natural or human-made incidents, such as ``Submarine cable disruption", government control or social instability, such as ``Egypt blackout" during the Egyptian revolution, or cyberattacks, such as ``Dyn cyberattack". Risk 38 ``Large-scale cyberattacks" includes ``Indiscriminate attacks", ``Destructive attacks", ``Cyberwarfare", ``Government espionage", ``Corporate espionage", ``Stolen e-mail addresses and login credentials", ``Stolen credit card and financial data", and ``Stolen medical-related data". Some of the cyberattacks are caused by geopolitical risks, such as ``Cyberattacks during the Russo-Georgian War". Some are related to social instabilities. Some are possibly caused by economic risks, such as ``2014 JPMorgan Chase data breach". It is also part of Risk 39 ``Massive incident of data fraud/theft", which includes major incidents of data breaches. One of the most significant data theft events is "Global surveillance disclosure" in 2013. Cyberattack and Data theft become significant during 2013 and 2014. Most of the other technological risks are unforeseen consequences of technological advances and may materialize in the future.

For a particular month, there are four different reasons to label a risk as being active. 
One is monthly statistical data, such as "unemployment rate", "oil price", "housing price", "deflation rate", and "food price index". They are recorded month by month, which can be directly used to label a risk. 
Another reason is yearly statistical data, such as "yearly deforestation rate", "annual greenhouse gas index", "corruption perceptions index", and "number of water conflicts per year". They are recorded year by year. Thus, we use one year’s single data point to estimate a risk status for twelve months. 
The third one is the record of daily events, such as "air travel disruption", "oil spill", "nuclear test", "terrorist attack", "blackout", and "cyber attacks". Within a certain month, if at least predefined number of events occurs, we consider the corresponding risk as active for the whole month.
The last reason is the record of continuous events, such as "European debt crisis", "subprime mortgage crisis", "hurricane seasons", "civil war", "spread of disease", "migration", and "data breaches". Since those events usually last several months, we consider corresponding risks as active from the beginning to the end of the event.

Of the five categories, economic and geopolitical risks have strong intra-dependence, environmental risks happens quite regularly and are relatively independent from each other, while societal and technological risks are affected by and have strong inter-dependence with geopolitical and economic risks.

\section*{Mean-field Steady State Points}
\subsection*{Simulation}

\begin{figure*}
	\centering
	\includegraphics[width=0.9\textwidth]{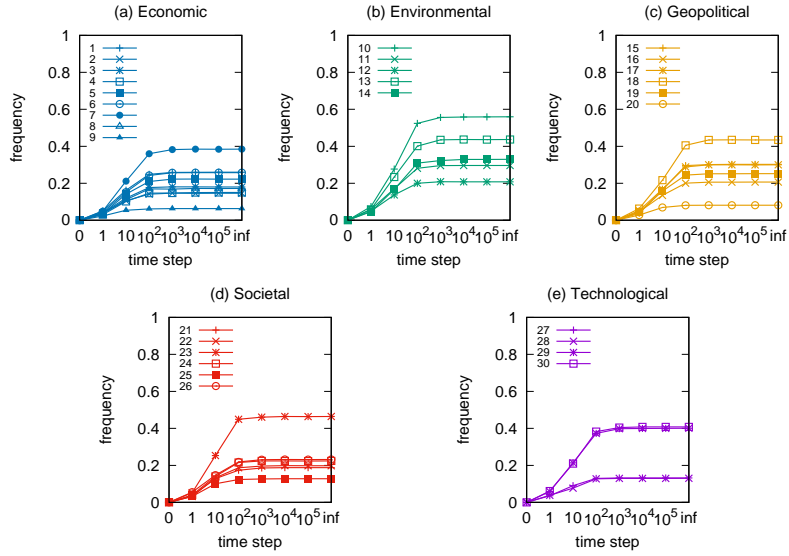}
	\caption{Asymptotic mean-field steady state frequencies of selected risks being active are shown for the 2017 risk network. Each frequency value is averaged over 1000 runs. For time denoted on the x-axis as "inf", the frequencies are calculated by Eq.~\ref{eq_risk_probability}. The environmental group of risks is most frequently active in this year, but the general level of risk activities is quite low.}
	\label{fig_cascade}
\end{figure*}

\begin{figure}
	\centering
	\includegraphics[width=1.2\textwidth, angle =-90]{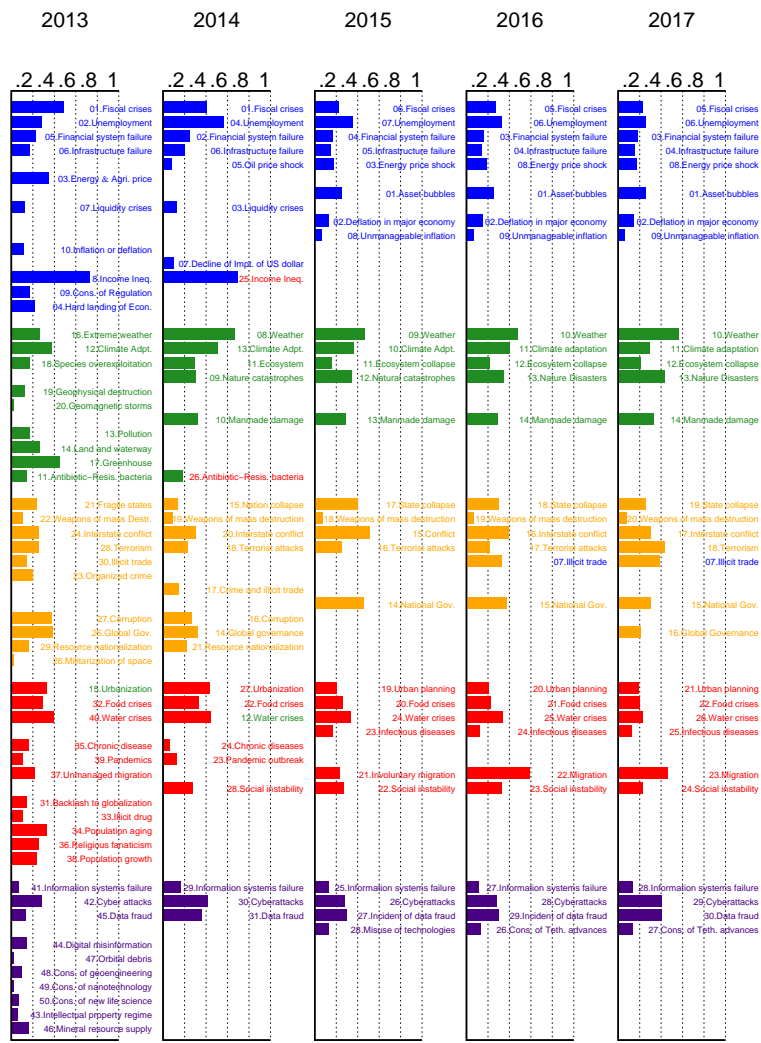}
	\caption{The mean-field steady state probabilities of risks being active for each network from 2013 to 2017. To see the changes in risks, we display related risks and their indices in the five networks side by side. In Table~\ref{table_risk_index} from 2013 to 2017 five risk categories remain the same, while around 20 risks vanish or merge into other risks. Visual inspection reveals that in 2013 the dominant risk was economical. It continued its dominance to 2014 but in that year the environmental group became also more active than the other groups. In year 2015 and in the following years, the economic risks drop their activity level, while environmental risk maintain significant level of activity in this period. In years 2016 social risks increase their activity levels, but weakened them in the following year.}
	\label{fig_steady_state}
\end{figure}

With the fitted parameters $\alpha, \beta, \gamma$, and the activation and recovery probabilities, we can perform Monte Carlo simulations of the cascades of global risks. Fig.~\ref{fig_cascade} shows the frequency of a risk being active at each time step $t$ during the simulation of the 2017 risk network with all risks initially inactive. With different initial states, the risk trajectories differ but eventually reach the same steady state. The frequency of risk $i$ being active at time $t$ is the ratio of the number of simulation months during which risk $i$ is active to the total number of simulation months $t$. The frequency distributions of risks being active mainly change from 10 to 1000 steps, and generally saturate afterward. In the steady state, the frequencies of risks being active varies a lot even for risks in the same category. By denoting the probability of risk $i$ being active at time $t$ as $p_i(t)$, we define such frequencies to be stable when $p_i(t) \approx p_i(t+1)$. By plugging in the state transition probabilities from Eq.~\ref{eq_transition}, we have
\begin{equation}
[1-p_i(t)]P_i(t)^{0\rightarrow1}+p_i(t)[1-P_i(t)^{1\rightarrow0}]=p_i(t+1)=p_i(t)  .\\
\label{eq_probability_transition}
\end{equation}
Thus, 
\begin{eqnarray}
\hat{p_i}&=&\frac{P_i^{0\rightarrow1}}{P_i^{0\rightarrow1}+P_i^{1\rightarrow0}}=\frac{1-(1-L_i)^{\alpha+\beta\sum_{j\in N_i}\hat{p_j}}}{1-(1-L_i)^{\alpha+\beta\sum_{j\in N_i}\hat{p_j}}+(1-L_i)^{\gamma}}   ,
\label{eq_risk_probability}
\end{eqnarray}
where $\hat{p_i}$ is the steady state  probability of risk $i$ being active, computed with a successive approximation method. The results are plotted in Fig.~\ref{fig_steady_state}.

\subsection*{Risk Evolution}

Fig.~\ref{fig_steady_state} shows the evolution of the global risks networks and their mean-field steady state points. To see the changes in risks, we display related risks and their indices in the five networks side by side. In Table~\ref{table_risk_index} from 2013 to 2017 five risk categories remain the same, while around 20 risks vanish or merge into other risks. Risk 05b "Extreme volatility in energy and agriculture prices" is changed to Risk 05a "Severe energy price shock" after 2014. The extreme volatility in agriculture prices is merged into Risk 27 "Food crises". Risk 07c "Unmanageable inflation or deflation" splits into risk 07a "Deflation in a major economy" and 07b "Unmanageable inflation". Risk 08 "Severe income disparity", 09 "Unforeseen negative consequences of regulation", and 10 "Hard landing of an emerging economy" are not discussed since/after 2014, because of the recovery of the global economy. Risk 14b "Unprecedented geophysical destruction" and 14c "Vulnerability to geomagnetic storms" are merged into 14a "Major natural catastrophes". Risk 15b "Irremediable pollution", 15c "Land and waterway use mismanagement" and 15d "Rising greenhouse gas emissions" are merged into 15a "Man-made environmental catastrophes". Risk 16 "Antibiotic-resistant bacteria" is not discussed after 2014. Risk 21a "Illicit trade" is changed from geopolitical to economic and not described in 2015. Risk 26 "Failure of urban planning" changes from an environmental to a geopolitical. Risk 30 "Large-scale involuntary migration" is not described in 2014. Risk 31 "Profound social instability" is newly proposed in 2014 while risks 32-36 are removed because of their low likelihoods and impacts. Finally, risks 40b-42 are excluded because most of them describe unforeseen consequences of advanced technologies and have not yet occurred.

\begin{figure}
	\centering
	\includegraphics[width=0.75\textwidth]{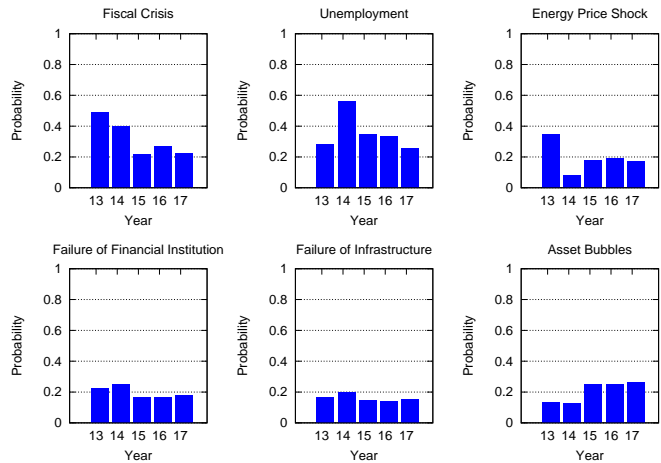}
	\caption{The mean-field steady state probabilities of economic risks being active for each of the networks from 2013 to 2017. Fiscal crisis significantly decreases from 2014, unemployment risk peaks in 2014, energy price shock peaks in 2013, while other risks vary minimally over the five years.}
	\label{fig_steady_state_economic}
\end{figure}

\begin{figure}
	\centering
	\includegraphics[width=0.75\textwidth]{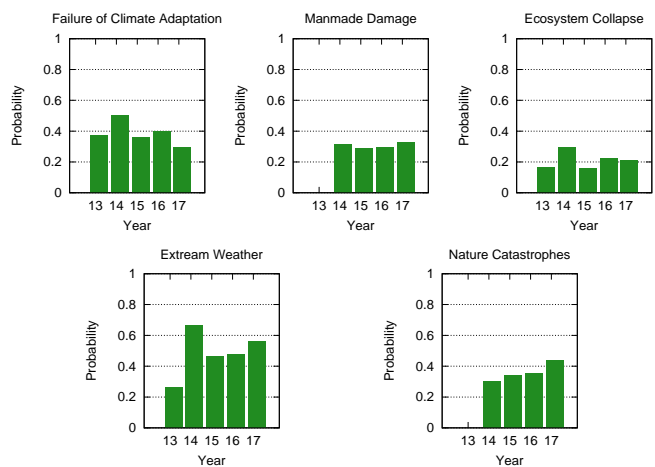}
	\caption{The mean-field steady state probabilities of environmental risks being active for each of the networks from 2013 to 2017. Man-made problems ("Failure of climate adaptation", "Ecosystem collapse") gradually decrease from 2013 to 2017, while natural disasters ("Extreme weather", "Natural catastrophes") gradually increase from 2013 to 2017.}
	\label{fig_steady_state_environmental}
\end{figure}

\begin{figure}
	\centering
	\includegraphics[width=0.75\textwidth]{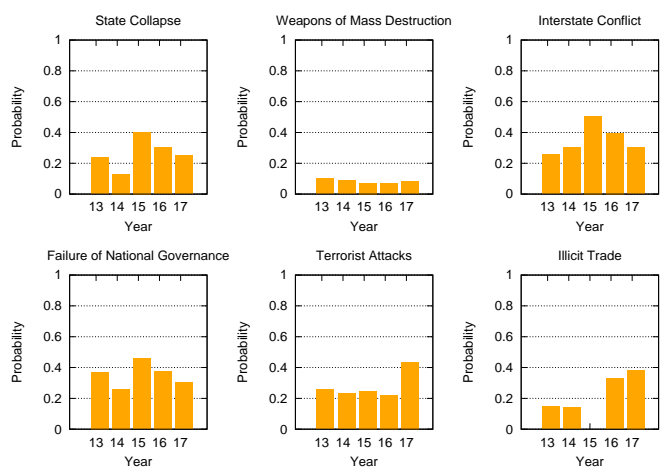}
	\caption{The mean-field steady state probabilities of geopolitical risks being active for each of the networks from 2013 to 2017.  "State collapse" and "Failure of national governance" have very similar behavior, since both of them describe risks inside nations and reflect the instability of their governments. Together with interstate conflicts, those risks significantly increase in 2015, and gradually decrease afterward. They may be caused by ISIS and the Ukraine crisis in 2014. Terrorist attacks happen more frequently in 2017, while growing illicit trade elevated levels of risk starting in 2016. This may be an affect of high risks of "State collapse", "Interstate conflict" and "National governance failure" during 2015. Risks associated with weapons of mass destruction remain low over the 5 year span.}
	\label{fig_steady_state_geopolitical}
\end{figure}

\begin{figure}
	\centering
	\includegraphics[width=0.75\textwidth]{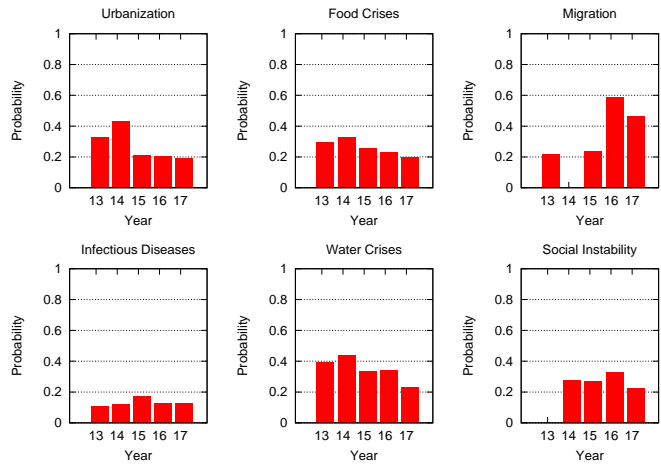}
	\caption{The mean-field steady state probabilities of societal risks being active for each network from 2013 to 2017. "Failure of urban planning" greatly decreases after 2014. Food and water crises gradually decrease with the global effort to address them. "Large-scale involuntary migration" drastically increases in 2016, due to the 2015 European Union migration crisis. It is largely affected by high risk of "State collapse", "Interstate conflict" and "National governance failure" in 2015. "Profound political and social instability" gradually increases from 2014 to 2016 as a consequence of rising "Interstate conflict", while "Rapid and massive spread of infectious diseases" maintains a steady and low risk level.}
	\label{fig_steady_state_societal}
\end{figure}

\begin{figure}
	\centering
	\includegraphics[width=0.75\textwidth]{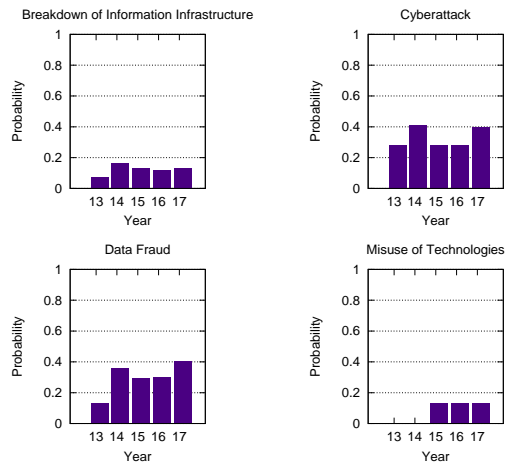}
	\caption{The mean-field steady state probabilities of technological risks being active for each network from 2013 to 2017. The peak of "Large-scale cyberattacks" and "Massive incident of data fraud/theft" in 2014 might be triggered by "Global surveillance disclosures" in the latter half of 2013. "Breakdown of critical information infrastructure and networks" and "Adverse consequences of technological advances" pose very little risk. Moreover, with fast-paced advancements in technology, all risks in this category gradually rise.}
	\label{fig_steady_state_technological}
\end{figure}

\begin{figure}
	\centering
	\includegraphics[width=0.75\textwidth]{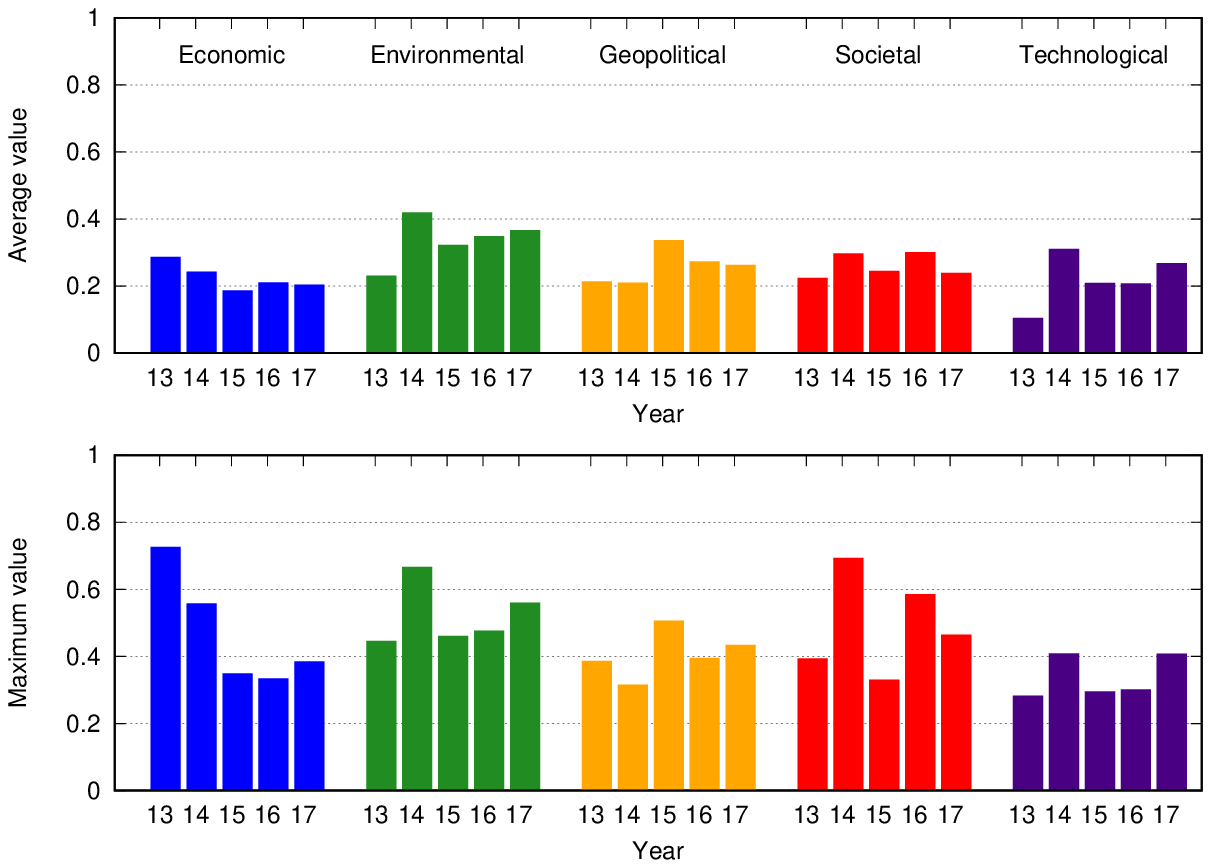}
	\caption{The mean-field steady state probabilities of risk activity in all categories for each of the networks from 2013 to 2017. In agreement with the results presented earlier, the economic risks drastically decrease from 2014 and maintain a low level afterward. Environmental risks occur regularly, with the exception of "Extreme weather" during 2014. The geopolitical risks have a small increase in 2015, during which interstate conflicts increase. The average values of societal risks experience a steady rise, but the maximum values have a high variance. That is because the Risk 08 "Severe income disparity" is put into societal risks in 2014, and because "Large-scale involuntary migration" drastically increases in 2016. Technological risks gradually increase with an anomalous spike encountered during 2014.}
	\label{fig_steady_state_category}
\end{figure}

Comparing the mean-field steady state probabilities of risks from 2013 to 2017 in Fig.~\ref{fig_steady_state_economic},~\ref{fig_steady_state_environmental},~\ref{fig_steady_state_geopolitical},~\ref{fig_steady_state_societal},~\ref{fig_steady_state_technological},~\ref{fig_steady_state_category}, we find that the probabilities of economic risks widely decreased, reflecting a gradual global recovery from the 2008 economic crisis. Only "State collapse", "Large-scale terrorist attacks", "Illicit trade" and "Large-scale involuntary migration" significantly increased in the category of geopolitical and societal risks, respectively. This reveals the downside of "Failure of global governance". Considering environmental risks, we find that man-made risks decreased, while only natural risks "Extreme weather events" and "Major natural disasters" increased. The ability of the public to prevent environmental degradation improved. In the technological risks category, "Massive incident of data fraud/theft" increases due to the boom of private data in the Internet era. The global risks transfer from economic to geopolitical and societal, as technological risks become heightened.

\subsection*{Analytic Probabilities and Empirical Observations}
Fig.~\ref{fig_steady_state} shows the analytic probabilities of risk being active at steady state over years. Fig.~\ref{fig_historical_events} depicts the time periods at which specific risk activities were observed and recorded. The analytic probabilities are smoother than the observations because they provide a real numbers instead of a binary signal. Empirical observations contain more detailed information than analytic probabilities. Although there are some differences, their main results are very consistent. Both results show the economic risks becoming inactive after year 2014. The environmental risks happen regularly every year. The geopolitical risks are highly interconnected. The risks ``State collapse", ``Interstate conflict" and ``Failure of national governance" have similar activity time scope. ``Terrorist attacks" and ``Illicit trade" are becoming important recently. The risk ``Migration" had active start from year 2016. The activity of technological risks gradually increases over years.

\section*{Model Validations}

\subsection*{Parameter Recovery Precision}
To test the accuracy of parameter recovery process for each year, we first take the learned $\alpha, \beta, \gamma$ as ground truth parameters, then use them to generate 125 test datasets with the same time steps of the historical data. From 125 test dataset, we learned 125 sets of new parameters that we use as sets of as test parameters. We consider activation parameter as $a\alpha+b\beta$, where $a, b$ is the average fraction of internal and external activation in a dataset. The recovery parameter equals to $\gamma$. From 125 sets test parameters, we first filter out 33.3\% outliers with the largest KS distance, defined by Eq. ~\ref{eq_ks_distance}, to the ground truth parameters, then determine the relative activation bound and recovery bound in the rest 66.6\% of the sets. We call those sets validation dataset and we refer to the corresponding parameters as validation parameters.
\begin{eqnarray}
KS(v_1, v_2) = \max_i(|v_1[i]/v_2[i]-1|)
\label{eq_ks_distance}
\end{eqnarray}
In this experiment, $v_1$ of Eq.~\ref{eq_ks_distance} is a vector of validation parameters, while $v_2$ is a vector of ground truth parameters. Each vector contains two variables, activation parameter and recovery parameter. Table~\ref{table_error_bound} shows the bounds of activation and recovery parameters for each year network. In 2013, recovery bound $0.087$ represents that within the validation parameters, the largest absolute relative error of recovery parameter to the ground truth recovery parameter is $0.087$. For all risk networks, the relative error bound of activation parameter is less than 20\%, the relative error bound of recovery parameter is around 10\%.

\begin{figure}
	\centering
	\includegraphics[width=\textwidth]{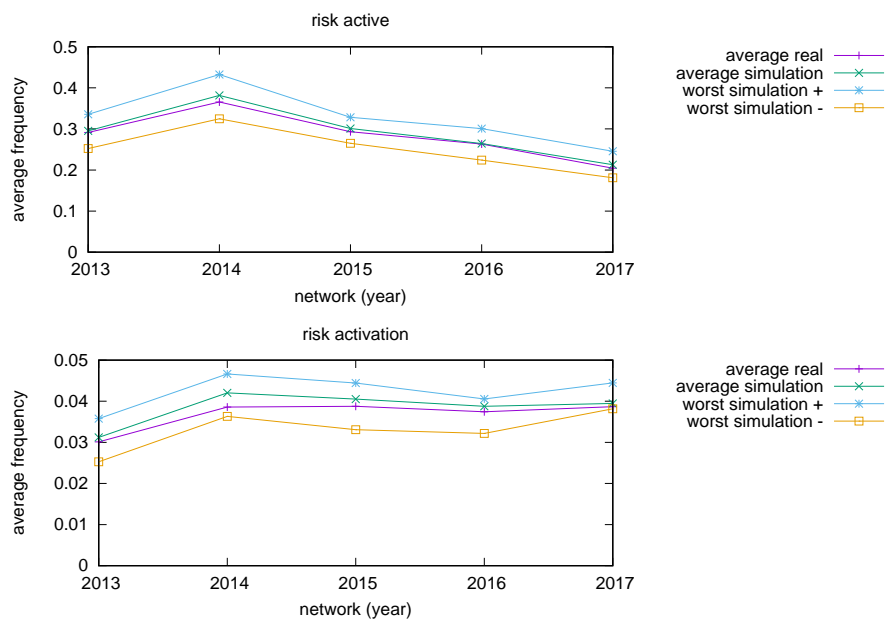}
	\caption{Error bounds of average frequencies of risk being active and of risk activation from the year 2013 to 2017. Average real represents the average results in the test of ground truth dataset. Average simulation represents the average results in the test of validation dataset. Worst simulation represents the maximum or minimum results in the test of validation dataset.}
	\label{fig_error_bound}
\end{figure}

Furthermore, we use the set of ground truth parameters and each set of validation parameters to generate another 12 months of data after the end of ground truth historical data. In this test, we run 100 realizations and calculate the average frequency of risk being active and risk activation for each set of parameters and the results are plotted in Fig.~\ref{fig_error_bound}. The average frequencies of risk being active are the sum of number of risks being active in each month averaged over $R$ risks and $12$ months. The average frequencies of risk activation are the number of times of any of the risks was activated over the entire simulation averaged over $R$ risks. In both tests of all risk networks, the average results of ground truth and validation data are very close. The absolute relative error of results in the worst simulation in validation data to the average results in ground truth data is around 20\%.

\begin{table}
\caption{Relative error bounds of activation and recovery parameters of global risks from the year 2013 to 2017. Clearly activation bounds are higher, up to 2.5 times higher than corresponding recovery bounds. The former peak at 20\% in 2016, while the latter peak at 12\% in 2015.}
\centering
\begin{tabular}{|c|c|c|c|c|c|} \hline
 & 2013 & 2014 & 2015 & 2016 & 2017 \\ \hline
activation bound & 0.181 & 0.125 & 0.174 & 0.200 & 0.188 \\ \hline
recovery bound & 0.087 & 0.104 & 0.119 & 0.077 & 0.071 \\ \hline
\end{tabular}
\label{table_error_bound}
\end{table}

\subsection*{Network Effects}
\begin{figure}
	\centering
	\includegraphics[width=\textwidth]{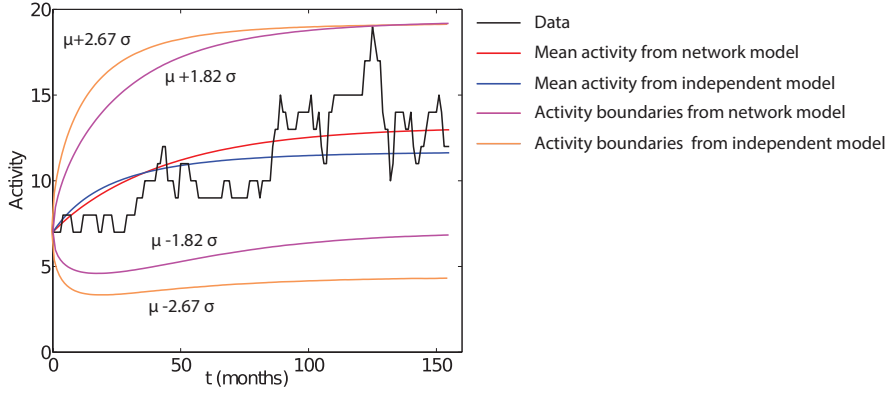}
	\caption{The average number of risk activations at each time step, measured over 100 runs, plotted as a function of time for the models with (red line) and without network effects (blue line). For comparison, the black curve represents the number of times risk activation was observed in the historical data. The purple and orange curves demonstrate what multiple of standard deviation is needed to get curves above and below the mean so the cover all points of historical data, for network model and independent model, respectively. This multiple is significantly lower, $1.82$, for network model than for independent model ($2.67$). Thus, the extreme historical data point among $156$ such points has probability of $3.4$\% to appear in network model, which is highly likely, but only $0.38$\% for the independent model, which in contrast is unlikely to appear with this number of historical data, demonstrating poor match between historical data and the independent model.}
	\label{fig_network_effect}
\end{figure}

We compare the simulation results by CARP model with and without network effect in Fig.~\ref{fig_network_effect}. This test is based on 2013 network and dataset from~\cite{szymanski2015failure}. The network model is the simulation of the 2013 network in Fig.~\ref{fig_network_evolution}, while the independent model ignores network effects by disregarding all edges. In general, compared with the independent model, the accuracy of the network model is significantly higher as evidenced by having the mean simulated activity closer to historical data than independent model does and by requiring 47\% smaller multiple of standard deviation bound to cover all historical data than the independent model needs. Some other network effect analyses were presented in~\cite{niu2017evolution}. The results show that the isolated risks (nodes with low degrees) have extremely low external activation fractions and thus are unlikely to be influenced by other risks in the network.

\subsection*{Sensitivity Tests}
\begin{figure}
	\centering
	\includegraphics[width=\textwidth]{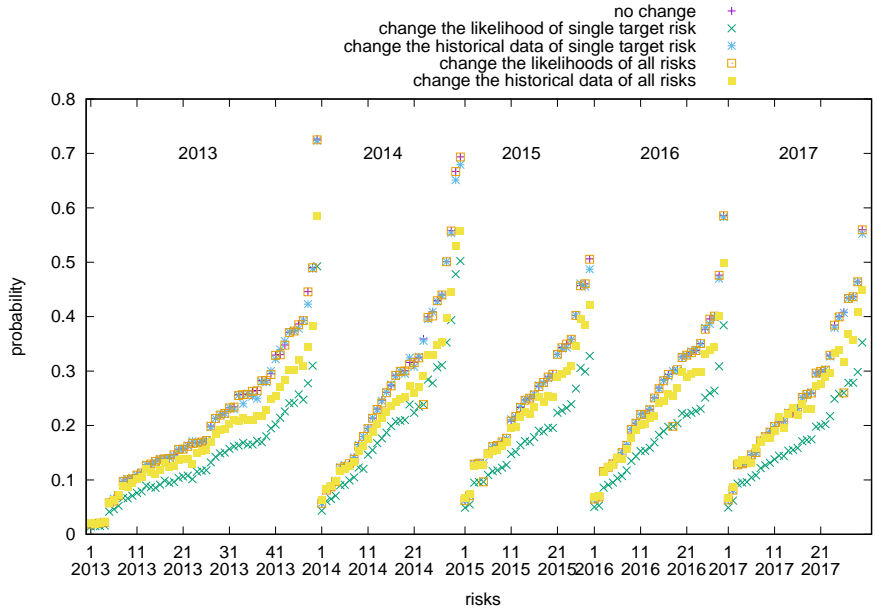}
	\caption{The comparison of the probability of risks being active at steady state by changing risk likelihood and historical data. The horizontal axis is the list of all risks grouped by different year and sorted by their probability. The baseline is the probability of risks without any change. The four other tests include reducing only: 1. the normalized likelihood of the single target risk by 10\%; 2. the frequency of the single target risk being active in the historical data by 10\%; 3. the normalized likelihoods of all risks by 10\%; 4. the frequencies of all risks being active in the historical data by 10\%.}
	\label{fig_sensitivity}
\end{figure}

In this subsection, we consider another two important factors: likelihood and historical data. Fig.~\ref{fig_sensitivity} shows the sensitivity test by changing the likelihood or historical data of single or all risks. The probability of a risk being active at steady state is sensitive to both likelihood and historical data. In the likelihood test, a risk is sensitive to the change of its single likelihood but tolerant to the change of all risks likelihoods. In the historical data test, on the contrary, a risk is tolerant to the change of its single historical activity but sensitive to the change of all risks historical activity.

\section*{External Activation}
\subsection*{Transition Fractions}
With the steady state probability of a risk being active, we can compute the probability of three different transition processes:
\begin{itemize}
	\item internal activation: $A_i^{int}=(1-\hat{p_i})p_i^{int}$ is the probability of inactive risk $i$ $(1-\hat{p_i})$ being triggered internally $p_i^{int}$.
	\item external activation: $A_i^{ext}=(1-\hat{p_i})[1-(1-p_{ji}^{ext})^{\sum_{j\in N_i}\hat{p_j}}]$ is the probability of inactive risk $i$ $(1-\hat{p_i})$ being triggered externally $1-(1-p_{ji}^{ext})^{\sum_{j\in N_i}\hat{p_j}}$.
	\item internal recovery: $A_i^{rec}=\hat{p_i}p_i^{rec}$ is the probability of active risk $i$ $(\hat{p_i})$ recovering $p_i^{rec}$.
\end{itemize}
For simplicity, we ignore the probability of a risk being activated both internally and externally with probability $(1-\hat{p_i})p_i^{int}[1-(1-p_{ji}^{ext})^{\sum_{j\in N_i}\hat{p_j}}]$ (the value is negligible). Thus, the three transition processes can be treated as independent variables. With the probabilities of transition processes, we can get the fraction of one transition process to all possible transitions for each risk by setting $a_i^{int}=\frac{A_i^{int}}{A_i^{int}+A_i^{ext}+A_i^{rec}}$, $a_i^{ext}=\frac{A_i^{ext}}{A_i^{int}+A_i^{ext}+A_i^{rec}}$, $a_i^{rec}=\frac{A_i^{rec}}{A_i^{int}+A_i^{ext}+A_i^{rec}}$. 

\subsection*{Risk Influence}
In this section, we calculate the influence exerted by one risk on others. In the experiments, we first disable a risk $i$ by setting its normalized likelihood $L_i=0$, and then calculate the new external activation frequency of risk $j$ as $a_{j-i}^{ext}$ ($j\neq i$). We obtain
\begin{equation}
I_{i\rightarrow j}=a_j^{ext}-a_{j-i}^{ext},
\end{equation}
where $I_{i\rightarrow j}$ is an indicator of the influence that risk $i$ exerts on risk $j$, quantifying the external activation effects of risk $i$ onto risk $j$.

\begin{figure}
	\centering
	\includegraphics[width=0.8\textwidth]{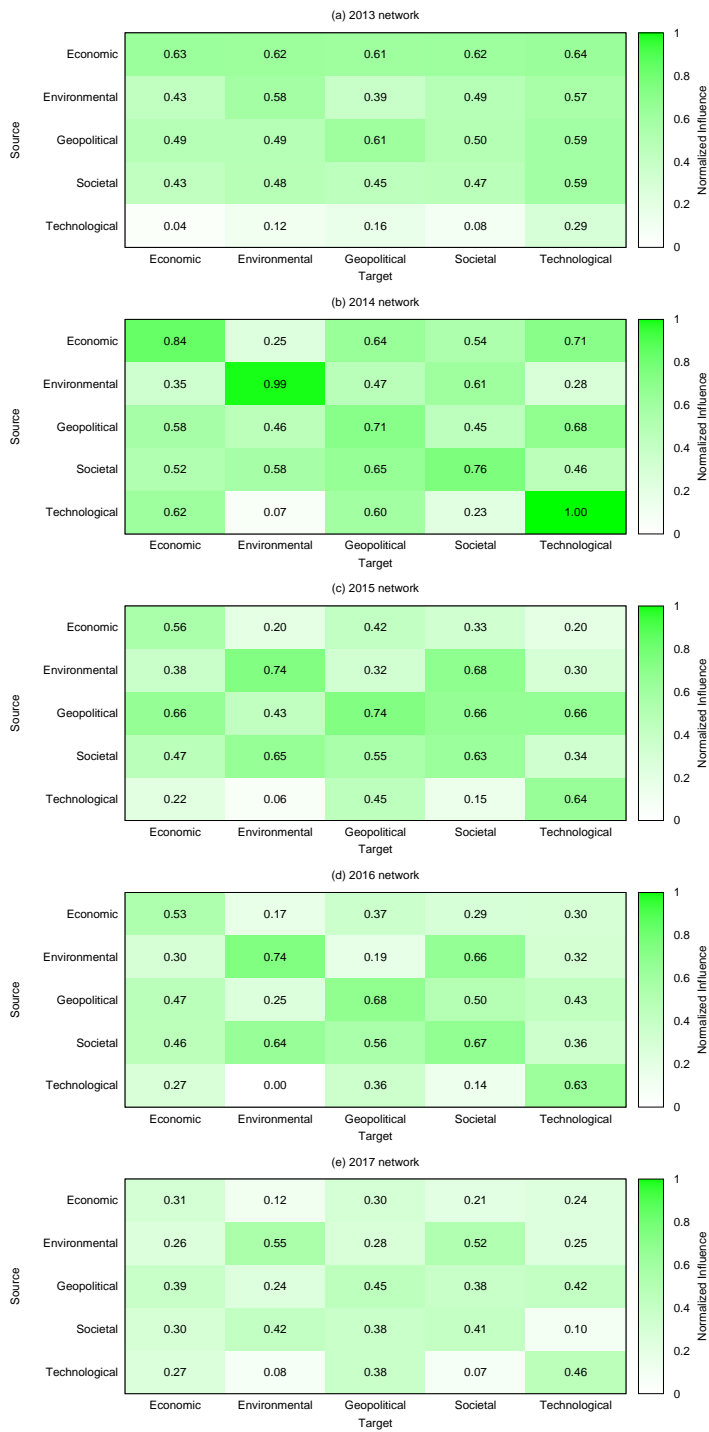}
	\caption{Influence among risk categories, normalized using a logarithmic scale. With the unity-based normalization of the influences, we find that most categories have large self-influence (diagonal elements).}
	\label{fig_category_influence}
\end{figure}

Fig.~\ref{fig_category_influence} shows the influence of a category of risks on other categories, which discerns between cause and correlation of risks. From 2013 to 2017, the most significant changes in risk influence categories are observed for economic and technological risks. The economic risks used to be the most influential risks and had the highest impact on other risks. However, as of 2017 their influence decreased. Instead, the influence of geopolitical and societal risks increased. In 2013, technological risks were the least vulnerable risks and had very limited influence on others. Although they are still the least influential risks in 2017, we can see an increasing trend in their influence. As shown in Fig.~\ref{fig_network_evolution} and~\ref{fig_network_property}, the 2014 and 2016 risk networks are sparser than the others. In a sparse risk network, risks have a higher tendency to connect with ones in the same category. Thus, the risk categories in 2014 and 2016 have largest self-influence in Fig.~\ref{fig_category_influence}.

\section*{Conclusions}
Here, we use the CARP model to simulate cascades in the global risk networks. With the most likely model parameters obtained through MLE (maximum likelihood estimation) and applied to a real event dataset, we compute the mean-field steady state probabilities of risks being active for each year from 2013 to 2017. The results obtained for the annual risk networks from 2013 to 2017 show significant changes in the asymptotic mean-field probabilities of risk activation. Applying the approach to finding bounds on recovery in CARP model presented in~\cite{lin2017limits} to the global risk network, we measure the error of model parameter recovery and find that it is bounded by $\pm$20\% of the values obtained with historical data for years from 2013 to 2017. The corresponding error of the risk activity is smaller but of similar magnitude. Since the range of values for critical risks reported above was much larger, we can conclude that we have enough historical data to support the conclusions of our paper. Finally, by computing the difference of external activation frequencies of risk $j$ with enabled and disabled risk $i$, we define the influence $I_{i\rightarrow j}$ that risk $i$ exerts on risk $j$. The results for the annual risk networks from 2013 to 2017 demonstrate that the influence among risks changes significantly over the years.

With the CARP model, we first compare yearly risks and then measure the quantitative changes of risks that provide an interesting view on evolution of the global economy and its risks. The activation probabilities and influences of economic risks are dramatically reduced as a result of economic recovery since 2014. The increase in activation probability of state collapse, terrorist attacks, illicit trade and migration show the negative effects of the failure of global governance, especially inaction of certain international bodies, like the Security Council of the United Nations. Technological risks are becoming more influential as well due to the increase of private data leaks. In each year from 2013 to 2017, the significance of economic threats decreases, while geopolitical and societal risks become more detrimental. All those analytic results are consistent with empirical observations. The quantitative analysis of our method creates a basis for developing tools for predictions of future risk network evolution and for guidance how to reduce damages caused by future risk cascades.

\section*{Abbreviations}
AGGI: Annual greenhouse gas index, CARP: Cascading alternating renewal processes, CTA: Collaborative Technology Alliance, DTRA: Defense Threat Reduction Agency, EU: European Union, FAO: Food and Agriculture Organization, ISIS: Islamic State in Iraq and Syria, KS: Kolmogorov-Smirnov, MLE: Maximum likelihood estimation, NATO: North Atlantic Treaty Organization, RPI: Rensselaer Polytechnic Institute, SIS: Susceptible-Infectious-Susceptible, U.S.: the United States, WEF: World Economic Forum

\section*{Availability of data and material}
Data are available online via WEF website, see bibliography.

\section*{Funding}
This work was supported in part by the Army Research Laboratory under Cooperative Agreement Number W911NF-09-2-0053 (the Network Science CTA), by the Army Research Office grant no. W911NF-16-1-0524, and by DTRA Award No. HDTRA1-09-1-0049. The views and conclusions contained in this document are those of the authors.

\section*{Competing interests}
The authors declare that they have no competing interests.

\section*{Author's contributions}
Designed research: X.N., B.K.S., G.K., A.M.; Performed research: X.N., B.K.S.; Analyzed data: X.N., B.K.S., G.K., A.M.; Wrote and edited the paper: X.N., A.M., G.K., B.K.S.

\section*{Acknowledgments}
The authors thank Dr. Noemi Derzsy for helpful discussion of this work in its preliminary stages. 

\section*{Author details}
$^1$Network Science and Technology Center, Rensselaer Polytechnic Institute (RPI), 110 Eighth Street, NY 12180 Troy, USA.
$^2$Department of Computer Science, Rensselaer Polytechnic Institute (RPI), 110 Eighth Street, NY 12180 Troy, USA.
$^3$Department of Physics, Rensselaer Polytechnic Institute (RPI), 110 Eighth Street, NY 12180 Troy, USA.

\bibliographystyle{bmc-mathphys} 
\bibliography{risk_cascade}

\end{document}